\documentclass[a4paper,11pt]{article}
\pdfoutput=1
\usepackage{jheppub} 
\usepackage{amsmath,amstext,amssymb}
\usepackage{comment,url}
\usepackage{graphicx,subfigure}
\usepackage{ulem}

\newcommand{\nn}{\nonumber\\}

\title{Momentum and charge transport in non-relativistic holographic fluids from Ho\v rava gravity}

\author[a]{Richard A. Davison,}
\author[b]{Sa\v so Grozdanov,}
\author[c]{Stefan Janiszewski,}
\author[d]{Matthias Kaminski}

\affiliation[a]{Department of Physics, Harvard University, \\ Cambridge, MA 02138, USA}
\affiliation[b]{Instituut-Lorentz for Theoretical Physics, Leiden University, \\ Niels Bohrweg 2,  Leiden 2333 CA, The Netherlands }
\affiliation[c]{Department of Physics and Astronomy, University of Victoria, \\ Victoria, BC, V8W 3P6, Canada}
\affiliation[d]{Department of Physics and Astronomy, University of Alabama, \\ Tuscaloosa, AL 35487, USA}

\emailAdd{rdavison@physics.harvard.edu}
\emailAdd{grozdanov@lorentz.leidenuniv.nl}
\emailAdd{stefanjj@uvic.ca}
\emailAdd{mski@ua.edu}

\abstract{We study the linearized transport of transverse momentum and charge in a conjectured field theory dual to a black brane solution of Ho\v rava gravity with Lifshitz exponent $z=1$. As expected from general hydrodynamic reasoning, we find that both of these quantities are diffusive over distance and time scales larger than the inverse temperature. We compute the diffusion constants and conductivities of transverse momentum and charge, as well the ratio of shear viscosity to entropy density, and find that they differ from their relativistic counterparts. To derive these results, we propose how the holographic dictionary should be modified to deal with the multiple horizons and differing propagation speeds of bulk excitations in Ho\v rava gravity. When possible, as a check on our methods and results, we use the covariant Einstein-Aether formulation of Ho\v rava gravity, along with field redefinitions, to re-derive our results from a relativistic bulk theory.}

\begin{document}
\maketitle
\flushbottom

\newpage

\section{Introduction}

The AdS/CFT correspondence~\cite{Maldacena:1997re} has proven to be an excellent tool with which to study the properties of certain strongly interacting, relativistic quantum field theories. It has taught us that these field theories have a robust hydrodynamic limit with a large window of applicability~\cite{Policastro:2002se,Policastro:2002tn,Bhattacharyya:2008jc,Grozdanov:2016vgg}, and has enabled the calculation of various hydrodynamic properties of these theories, most prominently the ratio of shear viscosity to entropy density \cite{Policastro:2001yc,Kovtun:2004de,Buchel:2004di,Buchel:2008sh,Myers:2008yi,Policastro:2002se,Kovtun:2005ev,Benincasa:2005qc,Baier:2007ix,Buchel:2008ac,Buchel:2008bz,Buchel:2008kd,Saremi:2011nh,Grozdanov:2014kva,Grozdanov:2015kqa,Kolekar:2016pnr}. For this reason, holography has provided a fertile testing ground for ideas about hydrodynamic descriptions of the quark-gluon plasma and of metals (see e.g.~\cite{CasalderreySolana:2011us,Hartnoll:2007ih}).

It is of both fundamental and practical interest to determine whether there are classical gravitational descriptions of strongly interacting field theories which are not relativistic at zero temperature. See \cite{Taylor:2015glc} for a recent review of this topic. One proposed class of gravitational duals are non-relativistic solutions of relativistic theories of gravity (general relativity (GR) coupled to appropriate matter content)~\cite{Son:2008ye,Balasubramanian:2008dm,Balasubramanian:2010uk}. A second approach~\cite{Janiszewski:2012nf,Janiszewski:2012nb}, which we pursue here, is to work with an intrinsically non-relativistic theory of gravity, like that proposed by Ho\v rava in~\cite{Horava:2009uw}. This theory is not invariant under all spacetime diffeomorphisms, and arises as the dynamical theory of Newton-Cartan geometry~\cite{Hartong:2015zia} (to which non-relativistic field theories naturally couple~\cite{Son:2013rqa,Jensen:2014aia,Hartong:2014pma,Fuini:2015yva}).

In this work, we study the linear response of a neutral black brane solution~\cite{Janiszewski:2014iaa} of (3+1)-dimensional Ho\v rava gravity. A manifestation of the non-relativistic nature of this state is that the low energy, linearized excitations of different fields propagate at different speeds, and each field has its own `sound horizon' (trapped surface) \cite{Blas:2011ni}. The causal `universal' horizon of the solution traps modes of arbitrarily high speed, and is the thermodynamic horizon of the solution~\cite{Blas:2011ni,Barausse:2011pu,Berglund:2012bu,Berglund:2012fk,Janiszewski:2014iaa,Bhattacharyya:2014kta}. The solution we study has an asymptotic Lifshitz symmetry with dynamical exponent $z=1$, i.e.~a scaling symmetry under which both time and space transform identically. It is invariant under spatial rotations and translations in space and time. However, this solution does not have a Lorentz boost symmetry, as this relativistic transformation is not a symmetry of Ho\v rava gravity. Moreover, the solution we study has no Galilean boost symmetry.

Using the proposed holographic dictionary of~\cite{Janiszewski:2012nf,Janiszewski:2012nb}, with further refinement following \cite{Jensen:2014aia,Jensen:2014wha,Hartong:2015zia}, we show that there is a simple hydrodynamic description of the linearized transport of both charge density and transverse momentum density over long times and distances in the conjectured dual field theory. In particular, both of these quantities diffuse, and have conductivities related to their diffusion constants by Einstein relations. This is an important consistency check of the existence of a holographically dual state of this black brane. In terms of bulk quantities, we find that the charge diffusion constant $D_\rho$ and the transverse momentum diffusion constant $D_\pi$ can be neatly expressed as
\begin{equation}
\label{eq:introdiffusionconstants}
\begin{aligned}
&D_{\rho} = (\text{speed}) \times (\text{sound horizon radius}),\\
&D_\pi = \frac{1}{3}\;(\text{speed}) \times (\text{sound horizon radius}), 
\end{aligned}
\end{equation}
where the relevant speed and sound horizon radius in each case is that of the corresponding dual excitation in the gravitational theory. We note that these constants have the same form as the analogous relativistic formulae, in which case different bulk excitations have the same speed and sound horizon radius due to Lorentz invariance. In terms of the temperature of the universal horizon, the diffusion constants are given by (\ref{eq:momdiffusionconstT}) and (\ref{eq:U1diffusionconstT}).

Our results for momentum transport are complementary to those of~\cite{Eling:2014saa} (which worked with a related, covariant theory\footnote{See Section \ref{sec:covarianthorava} for further comparison between these theories.}), in which a non-linear hydrodynamic description of transport was derived to leading order in perturbation theory in $\beta$, one of the coupling constants of Ho\v rava gravity. $\beta$ parameterizes the difference between the speed of one of the gravitons and the null speed of the boundary metric. While we study only linear (in amplitude) perturbations, we work non-perturbatively in $\beta$. Our non-perturbative result for the shear viscosity $\eta$ agrees with that conjectured in~\cite{Eling:2014saa}: $\eta/s=2^{2/3}/4\pi$, where $s$ is the entropy density. This does not match smoothly onto the GR result $1/4\pi$ when $\beta\rightarrow0$, as the universal horizon of the Ho\v rava solution does not coincide with the Killing horizon of the GR solution in the limit $\beta\rightarrow0$.

To obtain our results, we must modify the standard prescription for computing two-point retarded Green's functions in the relativistic case~\cite{Son:2002sd,Herzog:2002pc,Skenderis:2008dh,Skenderis:2008dg}, due to the existence of multiple horizons. We propose that the linear excitation of a field should obey ingoing boundary conditions at its sound horizon. In some cases, we are able to check that this is a sensible prescription by firstly rewriting Ho\v rava gravity in a covariant form (Einstein-Aether theory), and then using a field redefinition invariance of this theory, as well as diffeomorphisms, to map the perturbation equations onto those of the Schwarzschild-AdS$_4$ black brane solution of GR. This procedure maps the sound horizon radius of the original Ho\v rava solution to the Killing horizon radius of the Schwarzschild-AdS$_4$ black brane. It also provides a natural explanation for the appearance of the speeds of the bulk excitations, rather than the null speed of the boundary metric, in the diffusion constants (\ref{eq:introdiffusionconstants}). We expect that this general principle -- that ingoing boundary conditions should be applied at different values of $r$ for bulk excitations which travel at different speeds -- should be valid in Ho\v rava gravity beyond these simple cases.

To obtain finite answers for the transverse momentum density correlators, we performed holographic renormalization by including two counterterms which are invariant under the symmetries of Ho\v rava gravity. Upon the mapping to a covariant Einstein-Aether theory, these counterterms coincide with those of the GR calculation.

Finally, we exploit the field redefinition invariance of the covariant form of Ho\v rava gravity to identify a special point in the parameter space of the Ho\v rava theory (when the coupling $\lambda=0$), in which the full linear response dynamics of the black brane is equivalent to that of the Schwarzschild-AdS$_4$ solution of GR. Therefore, at this special point, the excitation spectrum of the dual field theory contains a sound mode (\ref{eq:sounddispersionrelation}) with speed proportional to the spin-$2$ graviton speed.

In the following section we provide a brief overview of linear response in hydrodynamics, and a derivation of the expected forms of the retarded Green's functions for charge density and transverse momentum density. In Sections \ref{sec:horavamomentumcalculation} and \ref{sec:horavachargecalculation} we study linear perturbations of the Ho\v rava black brane solution and derive from this the hydrodynamic forms of the dual Green's functions. The relation between our Ho\v rava gravity calculations and those of Einstein-Aether theory are described in Section \ref{sec:covarianthorava}, before we conclude in Section \ref{sec:discussion} with a summary of our results and some open questions.

\section{Hydrodynamics and linear response}
\label{sec:hydrosection}

In general, a system which is in local thermal equilibrium should have a coarse-grained, hydrodynamic description over long lengths and times, with respect to the scales over which the system is locally equilibrated (in our case, this is the inverse temperature). The hydrodynamic variables are those which vary slowly over these long length and time scales. These are typically the densities of the conserved charges of the system. 

We are interested in the linear response properties of a $U(1)$ charge density, and the transverse momentum density, in a (2+1)-dimensional, rotationally and translationally invariant field theory state with a $z=1$ scaling symmetry. We assume that the conserved charges of the state are its energy, $U(1)$ charge and momentum. The densities of these conserved charges $q_a$ obey the following conservation equations
\begin{equation}
\label{eq:hydrochargeconservationeqns}
\partial_t q_a+\vec{\nabla}\cdot\vec{j}_a=0,
\end{equation}
where $\vec{j}_a$ is the current density associated with the conserved charge density $q_a$. We will consider the response of states in which both the $U(1)$ charge density and the momentum density have vanishing expectation values. 

The information about the linear response properties of the state are contained in its two-point retarded Green's functions, which tell us how the expectation values of operators respond to small external sources. The retarded Green's functions of the charge densities and associated current densities can be computed within hydrodynamics using the canonical method of Kadanoff \& Martin \cite{1963AnPhy..24..419K} (see~\cite{Kovtun:2012rj} for a review). Heuristically, this method proceeds in two steps. When a small external source for a conserved charge density is applied at an initial time, the response in the expectation value of the charge density at that time is controlled by the susceptibility $\chi$.\footnote{Such susceptibilities are also sometimes referred to as `thermodynamic transport coefficients', e.g.~\cite{Jensen:2011xb}.} This initial change in the expectation value will then evolve in time via the equations of motion (\ref{eq:hydrochargeconservationeqns}), and the variation of this response at time $t$, with respect to the initial source, gives the retarded Green's function.

To determine the evolution in time of the charge densities, we must supplement the equations (\ref{eq:hydrochargeconservationeqns}) with constitutive relations for the current densities $\vec{j}_a$ in terms of the charges $q_b$. Hydrodynamics is a universal effective theory, and we therefore construct these relations by writing down all terms containing the conserved charges and their derivatives that are allowed by the symmetries of the system. The relations are written as a derivative expansion, and are a good approximation at long distance and time scales. The microscopic details of the system enter in the values of the coefficients of each term in these derivative expansions. After a Fourier transform in the spatial directions, and using the constitutive relations to replace $\vec j_a$ with $q_b$, the equations of motion for linear perturbations take the form
\begin{equation}
\partial_t q_a(t,k) + M_{ab}(k)q_b(t,k)=0.
\end{equation}
The two-point retarded Green's functions of the charges are then given by \cite{Kovtun:2012rj}
\begin{equation}
\label{eq:KMgeneralform}
G(\omega,k)=-\left(1+i\omega\left(-i\omega+M(k)\right)^{-1}\right)\chi.
\end{equation}

There has recently been a lot of progress in systematically constructing the full, non-linear constitutive relations of non-relativistic hydrodynamics~\cite{Kaminski:2013gca,Jensen:2014ama,Jensen:2014aia,Jensen:2014wha}, and also for Lifshitz hydrodynamics~\cite{Hoyos:2013eza,Hoyos:2013qna,Hoyos:2015yna,Hoyos:2015lra,Kiritsis:2015doa}. For our purposes this is overkill: by restricting to the linear reponse of parity-invariant theories, simple Kadanoff-Martin arguments are valid. It can be checked, for example, that imposing parity symmetry on the constitutive relations of \cite{Jensen:2014ama,Jensen:2014aia} reduces the constitutive relations written in terms of Newton-Cartan data to the usual Navier-Stokes equations.

Without loss of generality, we will align the $y$-axis with the direction along which linear perturbations vary in space. The conserved charge densities of our system are the energy density $\varepsilon$, the $U(1)$ charge density $\rho$, and the momentum densities $\pi_x$ and $\pi_y$. We also assume that parity is unbroken, and that charge conjugation, under which only the $U(1)$ charge density and current flip sign, is a symmetry of the state. This last condition implies we are studying a state which is not charged under this $U(1)$.

We begin with the constitutive relation for the longitudinal $U(1)$ charge current density $j_\rho^y$. The goal is to write down, to linear order in perturbations, the most general derivative expansion of the charges that is consistent with the symmetries above. To leading order in the derivative expansion, only one term is allowed
\begin{equation}
\label{eq:chargeconstrelation}
j_\rho^y=-D_\rho\nabla_y\rho+\ldots\,.
\end{equation}
The ellipsis denotes higher order terms in the derivative expansion. The constant $D_\rho$ is a transport coefficient that is not fixed by this analysis but depends upon microscopic details of the theory.

The linearized constitutive relation for the longitudinal current $j_{\pi_x}^y$ of the transverse momentum density $\pi_x$ is equally simple
\begin{equation}
\label{eq:momentumconstrelation}
j^y_{\pi_x}=-D_\pi\nabla_y\pi_x+\ldots\,.
\end{equation}
In this case, it is parity symmetry (under which $\pi_x\rightarrow-\pi_x$) that restricts the form of the right hand side. $D_\pi$ is a transport coefficient which, in general, is unrelated to $D_\rho$.

Combining the linearized constitutive relations with the conservation equations (\ref{eq:hydrochargeconservationeqns}), we find that linearized perturbations of both the charge density and the transverse momentum density obey a diffusion equation
\begin{equation}
\label{eq:hydrodiffusionequations}
\partial_t\rho-D_\rho\nabla^2\rho=0,\;\;\;\;\;\;\;\;\;\;\;\; \partial_t\pi_x-D_\pi\nabla^2\pi_x=0,
\end{equation}
and that the transport coefficients $D_\rho$ and $D_\pi$ are the diffusion constants of charge density and transverse momentum density, respectively. Diffusion constants have dimensions of $\text{speed}\times\text{distance}$.

From the diffusion equations (\ref{eq:hydrodiffusionequations}), we can use (\ref{eq:KMgeneralform}) to compute the hydrodynamic Green's functions of the conserved charges
\begin{equation}
\label{eq:hydrogreensfunctions}
G_{\rho\rho}(\omega,k)=\frac{\chi_\rho D_\rho k^2}{i\omega-D_\rho k^2},\;\;\;\;\;\;\; G_{\pi_x\pi_x}(\omega,k)=\frac{\chi_\pi D_\pi k^2}{i\omega-D_\pi k^2}.
\end{equation} 
Here, $\chi$ denote the static susceptibilities of the conserved charge densities
\begin{equation}
\left.\chi_\rho=\frac{\partial\rho}{\partial\mu}\right|_{\mu=0},\;\;\;\;\;\;\; \left.\chi_\pi=\frac{\partial\pi_x}{\partial v_x}\right|_{v_x=0},
\end{equation}
where the chemical potential $\mu$ is the source for the charge density, and the velocity $v_x$ is the source for the transverse momentum density. $\chi_\pi$ has units of mass density. The two-point retarded Green's functions of the associated current densities are fixed by Ward identities to be
\begin{equation}
\begin{aligned}
\label{eq:otherhydrogreensfunctions}
&G_{j_{\rho}^yj_\rho^y}(\omega,k)=\frac{\omega^2}{k^2}G_{\rho\rho}(\omega,k),\;\;\;\;\;\;\;\;\;\;\;\; G_{\rho j_\rho^y}(\omega,k)=G_{j_\rho^y\rho}(\omega,k)=\frac{\omega}{k}G_{\rho\rho}(\omega,k),\\
&G_{j_{\pi_x}^yj_{\pi_x}^y}(\omega,k)=\frac{\omega^2}{k^2}G_{\pi_x\pi_x}(\omega,k),\;\;\;\;\;\; G_{\pi_x j_{\pi_x}^y}(\omega,k)=G_{j_{\pi_x}^y\pi_x}(\omega,k)=\frac{\omega}{k}G_{\pi_x\pi_x}(\omega,k),
\end{aligned}
\end{equation}
up to contact terms.

In the long time (dc) limit, we define the linear response conductivities of $U(1)$ charge, and of transverse momentum as
\begin{equation}
\begin{aligned}
\label{eq:hydroeinsteinrelations}
&\sigma\equiv-\lim_{\omega\rightarrow0}\frac{1}{\omega}\text{Im}\left[\lim_{k\rightarrow0}G_{j_\rho^yj_\rho^y}(\omega,k)\right]=\chi_\rho D_\rho,\\
&\eta\equiv-\lim_{\omega\rightarrow0}\frac{1}{\omega}\text{Im}\left[\lim_{k\rightarrow0}G_{j_{\pi_x}^yj_{\pi_x}^y}(\omega,k)\right]=\chi_\pi D_\pi,\\
\end{aligned}
\end{equation}
respectively. The first of these corresponds to the usual definition of the conductivity via Ohm's law, and the second corresponds to the usual definition of the shear viscosity (see e.g.~\cite{zaanen-book}). The conductivities are fixed in terms of the diffusion constants by the Einstein relations (\ref{eq:hydroeinsteinrelations}), which follow simply from the form of the Green's functions (\ref{eq:hydrogreensfunctions}). From now on we will refer to these conductivities by their conventional names of the electrical conductivity and the shear viscosity, respectively.

We have refrained from a full discussion of non-relativistic~\cite{Jensen:2014ama} or Lifshitz~\cite{Hoyos:2013eza,Hoyos:2013qna,Hoyos:2015yna,Hoyos:2015lra} hydrodynamics and have presented only the elements which are relevant for our holographic computation. We have shown that transverse momentum and charge both diffuse, regardless of whether the system is relativistic or not. We note that the presence of an additional conserved particle number charge will not alter our conclusions, as it cannot enter the linearized constitutive relations (\ref{eq:chargeconstrelation}) and (\ref{eq:momentumconstrelation}) due to symmetry reasons. In the following sections, we will show that the Green's functions of the strongly interacting state purportedly dual to a Ho\v rava gravity black brane are of the hydrodynamic form (\ref{eq:hydrogreensfunctions}), and will derive explicit expressions for the transport coefficients $D_\pi$ and $D_\rho$ (or equivalently $\sigma$ and $\eta$) of this state.

\section{Momentum transport from a Ho\v rava black brane}
\label{sec:horavamomentumcalculation}

Ho\v rava gravity \cite{Horava:2009uw} is a non-relativistic quantum theory of gravity that breaks the local Lorentz covariance between space and time enjoyed by GR. We are interested in the low energy, classical regime of Ho\v rava gravity, whose degrees of freedom, $G_{IJ}$, $N^I$ and $N$, are the components of the ADM decomposition of a spacetime metric $g_{XY}$
\begin{equation}
g_{XY}dx^X dx^Y = -N^2dt^2+G_{IJ}\left(dx^I+N^Idt\right)\left(dx^J+N^Jdt\right).
\label{eq:adm}
\end{equation}
$G_{IJ}$ is the spatial metric on slices of constant global time $t$; $N$ is the lapse function, which encodes the normal distance between the leaves of the foliation by $t$; and $N^I$ is the shift vector, which identifies events with the same spatial coordinates on different time slices.\footnote{Our notation is that indices $X,Y\ldots$ are bulk spacetime indices with $x^0\equiv t$, while $I,J\ldots$ are bulk spatial indices covering both the bulk radial direction $x^1\equiv r$ and the transverse directions shared with the field theory $x^i\in (x,y)$.}

In (3+1)-dimensions, the two derivative bulk action of Ho\v rava gravity is
\begin{equation}
S^H=\frac{1}{16\pi G_H}\int d^4x\,N\sqrt{G}\left(K_{IJ}K^{IJ}-(1+\lambda)K^2+(1+\beta)(R-2\Lambda)+\alpha \frac{\nabla_I N\nabla^I N}{N^2}\right),
\label{eq:haction}
\end{equation}
where
\begin{align}
K_{IJ}\equiv \frac{1}{2N} \left(\partial_t G_{IJ}-\nabla_I N_J-\nabla_J N_I\right),
\end{align} 
is the extrinsic curvature of slices of constant $t$, $K$ is its trace, and $R$ and $G$ are the Ricci scalar and the determinant of the spatial metric, respectively. Indices are raised and lowered with $G^{IJ}$ and $G_{IJ}$, and $\nabla_I$ is the covariant derivative with respect to the spatial metric.

In addition to the cosmological constant $\Lambda$ and the gravitational constant $G_H$ (which has length dimension $2$), there are three new coupling constants $(\alpha,\beta,\lambda)$ allowed by the less restrictive symmetries of Ho\v rava gravity. These dimensionless constants must satisfy $\beta>-1$, $0\leq\alpha\leq 2(1+\beta)$, and $\lambda\geq 0$ or $\lambda\leq -2/3$, so that gravitons have positive speeds squared \cite{Griffin:2012qx}.

In comparison with the full spacetime diffeomorphism invariance of GR, Ho\v rava gravity is only invariant under the diffeomorphisms that preserve the foliation by slices of constant $t$. These are the spatial diffeomorphisms $x_I\to\tilde{x}_I(t,x_J)$ and reparametrizations of the global time $t\to \tilde{t}(t)$. In particular, spatially dependent time diffeomorphisms are not symmetries of Ho\v rava gravity.

\subsection{Ho\v rava black brane solution}

For the case $\alpha=0$, and with cosmological constant $\Lambda=-3$, there is an asymptotically AdS black brane solution to Ho\v rava gravity \cite{Janiszewski:2014iaa} with
\begin{align}
G_{IJ}&=\left( \begin{array}{c c c}
\frac{1}{r^2(1-\frac{r^3}{r_h^3})^2} & 0 & 0 \\
0& \frac{1}{r^2}& 0 \\
0 & 0& \frac{1}{r^2} \\
\end{array}\right),\;\;\;\;\;\;\;\;\;\;N=\frac{1}{r}\left(1-\frac{r^3}{r_h^3}\right),\;\;\;\;\;\;\;\;\;N_I=\left(\frac{r}{r_h^3}\frac{\sqrt{1+\beta}}{\left(1-\frac{r^3}{r_h^3}\right)},0,0\right).
\label{eq:solu}
\end{align}
We have chosen the AdS radius $L=1$, and used a radial coordinate $r$ which has an asymptotic boundary at $r=0$. This is a solution for values of $\lambda$ and $\beta$ consistent with the aforementioned constraints, and is smoothly connected to the numerical solutions of~\cite{Janiszewski:2014iaa}. We have checked (to leading order in $\alpha$) that, when $\alpha\ne0$, this solution has smooth corrections. The corresponding spacetime metric (\ref{eq:adm}) of this solution is asymptotically AdS, and the boundary metric has a ``null speed'' of 1. This is a choice of units, and all speeds in the formulae that follow are in units of this null speed.

The black brane solution (\ref{eq:solu}) has a ``universal horizon'' at $r=r_h$, where $N$, the normal distance between slices of constant $t$, vanishes \cite{Blas:2011ni,Barausse:2011pu,Berglund:2012bu,Berglund:2012fk,Bhattacharyya:2014kta}. In Ho\v rava gravity, causal signals propagate only forward in global time $t$. The leaves of the asymptotic temporal foliation that cover the boundary do not penetrate beyond $r=r_h$, where $N$ vanishes. Therefore, events at $r>r_h$ can only signal to larger $r$, and can have no causal influence on those at $r\leq r_h$. This causal event horizon traps modes of any speed, and is interpreted as the thermodynamic horizon of the solution \cite{Janiszewski:2014iaa}.

\begin{figure}
\begin{center}
\includegraphics[height=4cm]{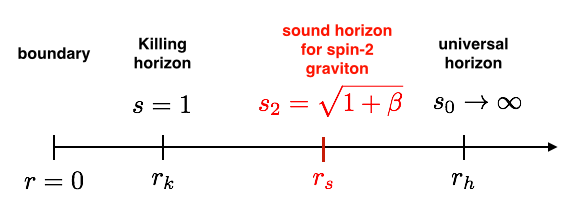}
\end{center}
\caption{\label{fig:HoravaHorizons}
An illustration of the horizons of the Ho\v rava black brane. The Killing horizon $r_k$, sound horizon of the spin-2 graviton $r_s$, and the universal horizon $r_h$, are trapped surfaces for waves with speed $s=1$, $s_2=\sqrt{1+\beta}$ and $s_0\rightarrow\infty$, respectively. Depending on the value of $\beta$, the spin-2 sound horizon $r_s$ can be inside or outside of the Killing horizon $r_k$.
}
\end{figure}

The Killing horizon of the solution is at $r_k\equiv r_h/(1+\sqrt{1+\beta})^{1/3}$. Its physical significance is that it is the trapped surface for modes of unit speed.\footnote{The locations of the various trapped surfaces, or sound horizons, for different speeds can be determined by examining the Killing horizon of an effective metric, as will be explicitly demonstrated in Section \ref{sec:covarianthorava}.} While the null speed of the asymptotic metric at the boundary is $1$, this is not the speed at which excitations of generic fields travel in Ho\v rava gravity. In contrast to GR, Ho\v rava gravity has more than one graviton. By examining the linearized field equations about the flat background $G_{IJ}=\delta_{IJ}$, $N_I=0$, and $N=1$, one finds a spin-2 graviton and an additional spin-0 graviton. The speeds squared of these modes are
\begin{align}
s^2_2=1+\beta,&& s^2_0=\frac{\lambda (1+\beta)}{\alpha(3\lambda+2)}\left(2\left(1+\beta\right)-\alpha\right),
\label{eq:speed}
\end{align}
respectively \cite{Griffin:2012qx}. Our background (\ref{eq:solu}) also supports multiple gravitons, and we will see shortly that the most important of these, for our purposes, travels at speed $s_2$. This is generically finite and therefore has a sound horizon (the trapped surface for modes of this speed) at a radius $r_s$, outside the universal horizon. When $\alpha=0$, $s_0\rightarrow\infty$, and a mode of this speed has a sound horizon which coincides with the universal horizon. A schematic location of the various horizons is shown in Figure \ref{fig:HoravaHorizons}.

\subsection{Black brane excitations}\label{sec:pert}

To determine the linear response of the conserved momentum density $\pi_x$ of our purported dual theory, we will study linearized excitations of the shift $\delta N_{x}(t,r,y)$ around the black brane solution. We will shortly outline in more detail the holographic dictionary that we use to explicitly identify the sources. $\delta N_{x}(t,r,y)$ couples to both $\delta G_{yx}(t,r,y)$ and $\delta G_{rx}(t,r,y)$. After choosing the gauge $\delta G_{rx}=0$ by making a radial diffeomorphism, the equations of motion for these linearized excitations are
\begin{align}
 \,k\sqrt{1+\beta}r_h^3r^3\left(\omega g(r)+kn(r)\right)+i(r^3-r_h^3)\left((1+\beta)k(2r^3-r_h^3)g'(r)-\omega r_h^3 n'(r)\right) &= 0,\nonumber\\
\,k r_h^6 r\left(\omega g(r)+k n(r)\right)+(r^3-r_h^3)\left(i k\sqrt{1+\beta} r^4g'(r)-(r^3-r_h^3)(-2n'(r)+r n''(r))\right) &= 0,\nonumber\\
 \,i r_h^3r\left(\sqrt{1+\beta}(2r^5+r_h^3r^2)+i\omega r_h^6 \right)(\omega g(r)+k n(r))&\nonumber\\
+2(r^3-r_h^3)\left((1+\beta)(r^6-r_h^6)-i\omega\sqrt{1+\beta} r_h^3 r^4\right)g'(r)&\nonumber \\
+r(r^3-r_h^3)\left(-i k\sqrt{1+\beta}r_h^3r^3n'(r)+(1+\beta)(2r^6-3r_h^3r^3+r_h^6)g''(r)\right) &= 0,
\label{eq:eomsshear}
\end{align}
where we have performed a Fourier transform with respect to the global time $t$ and the spatial direction $y$
\begin{equation}
\delta G_{yx}(t,r,y)\equiv\int d\omega dk e^{-i\omega t+iky}\frac{g(r)}{r^2},\;\;\;\;\;\;\;\;\;\;\;\;\;\;\delta N_x(t,r,y)\equiv\int d\omega dk e^{-i\omega t+iky}\frac{n(r)}{r^2}.
\end{equation}
We leave it implicit that $g$ and $n$ both depend upon $\omega$ and $k$.

Only one of the two second order equations of motion is linearly independent (because of the residual diffeomorphism invariance after our gauge choice  $\delta G_{rx}=0$), and we can make this manifest by working directly with the variable
\begin{align}
\psi(r)\equiv \frac{1}{r^2} \left(\omega g(r)+k n(r) \right) .
\label{eq:defnofgifield}
\end{align}
This field is invariant under the gauge freedom and obeys the second order equation
\begin{align}
\Big[q^4 z^2 (-2 + z^3)^2 (-1 + z^3) + 
 2 \nu^2 \left[-2 \nu^2 z^2 - i \nu z^4 (-5 + z^3) + (-2 + z^3)^2 (1 + 2 z^3)\right]&\nonumber\\
+q^2 \left[\nu^2 z^2 (8 - 8 z^3 + z^6) - 2 (2 - 3 z^3 + z^6)^2 + 
    i \nu z^4 (-10 + 6 z^3 + z^6)\right]\Big]\psi(z)&\nonumber\\
+z (-2 + z^3) \left[2 q^2 (-1 + z^3) (2 + z^3 (-3 - i \nu z + z^3)) \right. & \nonumber\\ 
   +\left. \nu^2 (4 + z^3 (-12 - 2 i \nu z + 5 z^3))\right]\psi'(z)&\nonumber\\
+z^2 (-2 + z^3)^2 (-1 + z^3) \left[\nu^2 + q^2 (-1 + z^3)\right]\psi''(z)&=0,			
\label{ap:eom}
\end{align}
where we have defined a rescaled radial coordinate $z$, frequency $\nu$ and wavenumber $q$ as
\begin{align}
z\equiv\frac{2^{1/3}r}{r_h},&& q\equiv \frac{r_h}{2^{1/3}}k,&& \nu\equiv \frac{r_h}{2^{1/3}\sqrt{1+\beta}}\omega.
\label{eq:rescale}
\end{align}
This rescaling manifestly removes $\beta$ from the equation of motion, as it entered only in the combination with $\omega$ that we have defined as $\nu$. The rescaled radial coordinateordinate is convenient as it has the sound horizon $r_s=r_h/2^{1/3}$ for the spin-2 perturbation at $z=1$. The universal horizon is at $z=2^{1/3}$ in these coordinates. Note that the linear response dynamics in this sector are completely independent of the coupling constant $\lambda$ of Ho\v rava gravity.

\subsection{Holographic dictionary}

To extract the linear response correlators of the dual field theory, we will follow a similar procedure as in the relativistic case, e.g. \cite{Policastro:2002se,Son:2002sd}. Firstly, we must solve the differential equation ({\ref{ap:eom}) subject to two boundary conditions. Our first boundary condition is to fix the overall normalization of the solution by demanding that $g(z=0)=g_0(\omega,k)$ and $n(z=0)=n_0(\omega,k)$, with $g_0$ and $n_0$ holographic sources for the dual field theory operators. 

To identify the sources in Ho\v rava gravity, we will use a refined holographic dictionary first presented in \cite{Janiszewski:2012nb,Janiszewski:2012nf}. This was originally motivated by the understanding of non-relativistic symmetry groups of \cite{2006AnPhy.321..197S,Son:2008ye}, and now has a more rigorous formulation in terms of the Newton-Cartan geometry of \cite{Jensen:2014aia,Hartong:2015zia}. The relation to Newton-Cartan geometry is most clearly illustrated by comparing the large $c\to\infty$ limits of \cite{Janiszewski:2012nb} and \cite{Jensen:2014wha}. In \cite{Jensen:2014wha}, the most general spacetime metric is written as
\begin{eqnarray}
g_{XY}\equiv -n_Xn_Y+h_{XY},
\label{eq:ncexpansion}
\end{eqnarray}
which is satisfied by the ADM decomposition (\ref{eq:adm}) for\footnote{To make this identification unambiguous, powers of $c$ need to be reinstated in the ADM expansion, as in \cite{Janiszewski:2012nb}.}
\begin{eqnarray}
n_X=(-N,0),\quad h_{XY}=\left(\begin{array}{c c}
N^KN_K & N_J \\
N_I & G_{IJ}
\end{array}\right).
\label{eq:ncadm}
\end{eqnarray}
In addition to this timelike vector $n_X$ and the degenerate symmetric ``metric'' $h_{XY}$, Newton-Cartan geometry contains a ``velocity'' field $v^X$ and an ``inverse metric'' $h^{XY}$ that are defined to obey
\begin{eqnarray}
h_{XY}v^Y=0,\quad n_Xv^X=1,\quad h^{XY}n_Y=0,\quad h^{XY}h_{YZ}=\delta^X_Z-v^Xn_Z,
\label{eq:vinv}
\end{eqnarray} 
which implies that they can be expressed in terms of the ADM fields as
\begin{eqnarray}
v^X=\left(-\frac{1}{N},\frac{N^I}{N}\right),\quad h^{XY}=\left(\begin{array}{c c}
0 & 0 \\
0 & G^{IJ}	
\end{array}\right).
\label{eq:vinvadm}
\end{eqnarray}
We are now in position to express the sources of the field theory in terms of the boundary values of the Ho\v rava fields by using the definition of Newton-Cartan sources found in \cite{Jensen:2014aia} (see also \cite{Hartong:2014pma}): $n_0$ is the source for energy density,\footnote{The lack of spatial components $n_i$ renders us unable to calculate the energy current.} $\bar h^{\mu\nu}$ is the source for the stress tensor, and $\bar v^\mu$ is the source for momentum density. The barred notation is due to the fact that the sources should be varied arbitrarily, while the fields $v^\mu$ and $h^{\mu\nu}$ must obey the constraints (\ref{eq:vinv}). The explicit relation between variations of barred sources and unbarred fields are given in \cite{Jensen:2014aia}, but for our background they are expressed in terms of the bulk fields as
\begin{eqnarray}
\delta n_0=-r\delta N|_{r=0},\quad \delta \bar h^{ij}=\frac{1}{r^2}\delta G^{ij}|_{r=0},\quad \delta\bar v^i=\delta N^i|_{r=0}, 
\label{eq:dict}
\end{eqnarray}
where the powers of $r$ are needed to strip off the leading behavior of the bulk fields as we approach the boundary at $r\to 0$. The sources of stress $\Pi_{yx}$ and momentum density $\pi_x$ are therefore
\begin{eqnarray}
\delta \bar h^{yx}=\frac{1}{r^2}\delta G^{yx}|_{r=0}=r^2\delta G_{yx}|_{r=0}\sim g_0, \quad \delta\bar v^x=\delta N^x|_{r=0}=r^2\delta N_x|_{r=0}\sim n_0,
\label{eq:sources}
\end{eqnarray}
respectively. This agrees with the discussion of boundary conditions at $r\to 0$ above. In the notation of Section \ref{sec:hydrosection}, $\Pi_{yx}$ is equal $j^y_{\pi_x}$, the longitudinal component of the current associated with the conserved charge density $\pi_x$.

Now that we understand the boundary condition at $r=0$, giving the normalization of the bulk fields in terms of field theory sources, we need to apply another boundary condition in order to solve the second order equation of motion (\ref{ap:eom}). In the relativistic case, to determine the retarded Green's function of the dual field theory, one must impose ingoing boundary conditions at the black brane horizon \cite{Policastro:2002se,Son:2002sd}. Heuristically, this is because the retarded Green's function is the causal response function in the field theory, and causality in the bulk implies that nothing should come out of the black hole. The situation is more subtle in Ho\v rava gravity; we must take care as there are multiple horizons. In fact, the equation of motion (\ref{ap:eom}) has singular points at both the spin-2 sound horizon and the universal horizon. By studying the characteristic exponents near each singular point, we find that it is only possible to impose ingoing boundary conditions at the spin-2 horizon, which is the outermost singular point. We therefore choose this location to impose ingoing (in global time) boundary conditions. These boundary conditions, as well as the identification of sources, will be further justified in Section \ref{sec:covarianthorava} via a mapping to a covariant calculation.

After imposing these boundary conditions, we will determine the dual Green's functions from the on-shell action of Ho\v rava gravity, as in the relativistic case. This step requires an appropriate holographic renormalization to obtain a finite answer, as will be explained shortly.

\subsection{Hydrodynamic solution of the equation of motion}

The equation of motion (\ref{ap:eom}) cannot be solved analytically in general. It can be solved analytically in a perturbative expansion at small frequencies and wavenumbers. This is sufficient for our purposes as we are ultimately interested in the dual Green's function in this hydrodynamic limit. Anticipating the existence of a diffusive excitation, we will perform a perturbative expansion in small $\epsilon$ where $q\sim\epsilon$ and $\nu\sim\epsilon^2$.

As explained above, we first impose ingoing boundary conditions at the spin-2 horizon $z=1$
\begin{equation}
\psi(z)=\left(1-z\right)^{-2i\nu/3}\frac{1}{z^2}\delta\psi(z),
\end{equation}
where $\delta\psi(z)$ is regular at $z=1$. The extra factor of $z^{-2}$ is chosen such that $\delta\psi\sim z^0$ as $z\rightarrow0$. To find the perturbative solutions, we expand
\begin{equation}
\delta\psi(z)= \epsilon \delta\psi_1(z)+\epsilon^2\delta\psi_2(z)+\epsilon^3\delta\psi_3(z)+\ldots,
\end{equation}
and solve order by order in $\epsilon$, imposing regularity at the horizon of $\delta\psi(z=1)$ at each order. At leading order in the small $\epsilon$ hydrodynamic expansion, the solution is
\begin{equation}
\delta\psi_1(z)=C_1+C_2z^3.
\end{equation}
We can identify these constants in terms of the near-boundary expansions of the fundamental fields
\begin{align}
\label{eq:nearbdyexpansions}
g(r\rightarrow0)&=g_0(\epsilon)+g_1(\epsilon)r+g_2(\epsilon)r^2+g_3(\epsilon)r^3+\ldots,\\
n(r\rightarrow0)&=n_0(\epsilon)+n_1(\epsilon)r+n_2(\epsilon)r^2+n_3(\epsilon)r^3+\ldots,
\end{align}
using the definition (\ref{eq:defnofgifield}). Note that while the normalization of these fields is given by the sources $g_0$ and $n_0$, and that $g_1$, $g_2$, $n_1$, and $n_2$ are determined in terms of these sources by the near boundary equations of motion, $g_3$ and $n_3$ are unfixed by near boundary analysis, and are related to the expectation values of the fields dual to $g$ and $n$. Comparing the expansion (\ref{eq:nearbdyexpansions}) to the leading order hydrodynamic solution yields
\begin{equation}
\delta\psi_1(z)=\frac{2q}{r_h^3}n_0(0)+qn_3(0)z^3.
\end{equation}

Moving to the first subleading order in the perturbative expansion, $\delta\psi_2$ obeys the same equation of motion as $\delta\psi_1$. Identifying the constants in terms of the near-boundary expansions of the fundamental fields, we find that
\begin{equation}
\delta\psi_2(z)=\frac{2}{r_h^3}\left(\sqrt{1+\beta}\nu g_0(0)+qn_0'(0)\right)+z^3\left(\sqrt{1+\beta}\nu g_3(0)+qn_3'(0)\right).
\end{equation}
Recall that in addition to the second order equation we are solving for $\psi$, there is also a first-order constraint equation (\ref{eq:eomsshear}) for $g$ and $n$. This constraint equation places further restrictions on the allowed near-boundary expansions (\ref{eq:nearbdyexpansions}). In particular it requires that
\begin{align}
g_3(0)=0, && g_3'(0)=-\frac{\nu}{q}\frac{1}{\sqrt{1+\beta}}n_3(0).
\end{align}

To determine the Green's functions to leading order, we do not need to explicitly solve the equations of motion at higher order in $\epsilon$. However, we do need to identify the subleading coefficients $n_3(0)$ and $n_3'(0)$ in terms of the boundary values of the fields. To do this, it is sufficient to expand the $O(\epsilon^3)$ and $O(\epsilon^4)$ equations of motion around $z=1$, and demand regularity of $\delta\psi_3(z)$ and $\delta\psi_4(z)$ here. This results in
\begin{align}
n_3(0)=\frac{2q^2n_0(0)}{r_h^3(3i\nu-q^2)},&& n_3'(0)=\frac{2\sqrt{1+\beta}q\nu g_0(0)+2q^2n_0'(0)}{r_h^3(3i\nu-q^2)}.
\label{eq:vevs}
\end{align}

\subsection{Holographic renormalization and hydrodynamic Green's functions}

The final step in our calculation of the retarded Green's functions in the hydrodynamic limit is to evaluate the on-shell gravitational action at the boundary $r\rightarrow0$. As in the usual relativistic case, we must supplement the action (\ref{eq:haction}) with two kinds of terms to obtain the correct answer. Firstly, we have to add a Gibbons-Hawking-like boundary term such that the variation problem is well-defined. As shown in \cite{Janiszewski:2014iaa}, the Ho\v rava-Gibbons-Hawking term only contributes to time-like boundaries, such as the asymptotic boundary $r=0$ in the current case. It can be written as:
\begin{equation}
S^{HGH}=\frac{1+\beta}{8\pi G_H}\int dt dx dy N \sqrt{H}\left(\,^2\mathcal{K}\right)\big|_\text{boundaries},
\label{eq:hgh}
\end{equation}
where $H$ is the determinant of $H_{ij}$ (the induced spatial metric of the boundary), and $^2\mathcal{K}$ is the trace of its extrinsic curvature, as embedded in the bulk slices of constant $t$. Note that $^2\mathcal{K}=\nabla_I s^I$ where $s^I$ is the unit spatial vector normal to surfaces of constant $r$. 

Secondly, we must renormalize the boundary action so that it is finite. To do this, we supplement the action with counterterms of boundary geometric objects that are invariant under foliation-preserving diffeomorphisms (the symmetries of the theory). A finite answer is obtained by including the following two counterterms
\begin{align}
&S^{HCT,1}= \frac{1+\beta}{4\pi G_H}\int dt dx dy N\sqrt{H},\\
&S^{HCT,2}=\frac{1}{16\pi G_H}\int dt dx dy N\sqrt{H}\left(\,^2K_{ij}\,^2K^{ij}\right),
\label{eq:hcts}
\end{align}
where $\,^2K_{ij}$ is the extrinsic curvature of the slices of constant $t$, as embedded in the boundary spacetime. 

The result of including all of these terms is that the on-shell quadratic action is now manifestly finite, and is of the form
\begin{align}
S^{H,\text{Total}}_2\Big|_\text{on-shell}=&\, \int d\omega dk\, \Big[ g_{0}(-\omega,-k)\mathcal{G}_{gg}(\omega,k)g_{0}(\omega,k)+g_{0}(-\omega,-k)\mathcal{G}_{gn}(\omega,k)n_{0}(\omega,k)  \nonumber\\
&+n_{0}(-\omega,-k)\mathcal{G}_{ng}(\omega,k)g_0(\omega,k)+n_{0}(-\omega,-k)\mathcal{G}_{nn}(\omega,k)n_{0}(\omega,k) \Big ],
\label{eq:onshellshear}
\end{align}
where we have explicitly reinstated the dependence upon $\omega$ and $k$. Equipped with the on-shell action, we now determine the retarded Green's functions of the operators $\pi_x$ and $j^y_{\pi_x}$ using the standard relativistic prescription. This prescription is that the retarded Green's functions $G_{AB}(\omega,k)$ are given by
\begin{align}
G_{AB}(\omega,k)=2\mathcal{G}_{\varphi_A\varphi_B}(\omega,k), 
\label{eq:gfprescription}
\end{align}
where $\varphi_A$ is dual to the operator $A$. We identified the fields dual to $\pi_x$ and $j^y_{\pi_x}$ ($\Pi_{yx}$) in (\ref{eq:sources}) as $n$ and $g$, respectively. One can roughly think of (\ref{eq:gfprescription}) as varying a generating functional, provided by the on-shell gravity action, with respect to the sources $n_0$ and $g_0$.

\subsection{Transport coefficients and susceptibilities}
\label{sec:momentumtransportGFresults}
Using the solution of the linearized equation of motion in the small $\omega$ and $k$ limit, and the prescription (\ref{eq:gfprescription}), the hydrodynamic retarded Green's functions of the operators in the dual field theory are (up to contact terms independent of $\omega$ and $k$)
\begin{align}
G_{\pi_x,\pi_x}(\omega,k)&=\frac{\sqrt{1+\beta}\,k^2}{8\pi G_H 2^{1/3}r_h^2\left(i\omega-\frac{\sqrt{1+\beta}r_hk^2}{2^{1/3}3}\right)},\\
G_{j_{\pi_y}^x,\pi_x}(\omega,k)&=\frac{\sqrt{1+\beta}\,k\omega}{8\pi G_H 2^{1/3}r_h^2\left(i\omega-\frac{\sqrt{1+\beta}r_hk^2}{2^{1/3}3}\right)},\\
G_{\pi_x,j_{\pi_y}^x}(\omega,k)&=G_{j_{\pi_y}^x,\pi_x}(\omega,k),\\
G_{j_{\pi_y}^x,j_{\pi_y}^x}(\omega,k)&=\frac{\sqrt{1+\beta}\,\omega^2}{8\pi G_H 2^{1/3}r_h^2\left(i\omega-\frac{\sqrt{1+\beta}r_hk^2}{2^{1/3}3}\right)}.
\end{align}

These Green's functions have the characteristic form associated with diffusive transport, in agreement with that expected based upon hydrodynamic considerations. The hydrodynamic Green's functions (\ref{eq:hydrogreensfunctions}) and (\ref{eq:otherhydrogreensfunctions}) have two free parameters that are determined by microscopic details of the specific theory: the momentum susceptibility $\chi_\pi$ and the momentum diffusion constant $D_\pi$. For the field theory purportedly dual to our solution of Ho\v rava gravity, these take the values
\begin{equation}
\label{eq:holodiffconstant}
\chi_\pi=\frac{3}{8\pi G_H}r_h^{-3},\;\;\;\;\;\;\;\;\;\;\;\;\;\;\;\;\;\;\;\; D_\pi=\frac{\sqrt{1+\beta}}{3\cdot 2^{1/3}}r_h.
\end{equation}
Our Green's functions obey the Einstein relation (\ref{eq:hydroeinsteinrelations}) and thus the field theory viscosity (or conductivity of transverse momentum) is
\begin{equation}
\label{eq:holoviscosity}
\eta=\frac{\sqrt{1+\beta}}{8\pi G_H \cdot 2^{1/3}}r_h^{-2}.
\end{equation}
The diffusion constant agrees with the expression (\ref{eq:introdiffusionconstants}) given in the introduction.

The Ho\v rava black brane solution (\ref{eq:solu}) obeys a first law-like relation \cite{Janiszewski:2014iaa}
\begin{equation}
d\epsilon=Tds,
\end{equation}
where $\epsilon$ is the ADM energy density of the solution, and the entropy density $s$ and temperature $T$ are properties of the universal horizon
\begin{align}
\label{eq:holothermodynamics}
\epsilon=\frac{1+\beta}{4\pi G_H r_h^3},&&s=\frac{\sqrt{1+\beta}}{4G_H r_h^2},&& T=\frac{3\sqrt{1+\beta}}{2\pi r_h}.
\end{align}
In terms of these thermodynamic quantities, the momentum susceptibility is
\begin{equation}
\label{eq:holomomsusc}
\chi_\pi=\frac{3\epsilon}{2\left(1+\beta\right)}=\frac{\epsilon+p}{1+\beta},
\end{equation}
where we have identified the pressure $p=\epsilon/2$ due to the $z=1$ scaling symmetry. Recall that $\chi_\pi$ should have units of mass density. The expression (\ref{eq:holomomsusc}) is telling us that the relevant speed for turning the energy density into a mass density is in fact $s_2=\sqrt{1+\beta}$, rather than the null speed of the boundary metric.

In the units $\hbar=k_B=1$ that we are using, diffusion constants have dimensions of speed$^2/$energy. Expressing the diffusion constant (\ref{eq:holodiffconstant}) in terms of $T$, the natural measure of energy of the fluid, we find
\begin{equation}
\label{eq:momdiffusionconstT}
D_\pi=\frac{1+\beta}{2^{4/3}\pi T},
\end{equation}
where it is again apparent that $s_2=\sqrt{1+\beta}$ is the natural speed of the system. 

The ratio of viscosity to entropy density of the system takes the $\beta$-independent value
\begin{equation}
\label{eq:etaovershorava}
\frac{\eta}{s}=\frac{2^{2/3}}{4\pi}.
\end{equation}
This result was conjectured in \cite{Eling:2014saa}, based upon a perturbative calculation to leading order in $\beta$, and we have provided an explicit verification of it. In light of the evidence above that $\beta$ appears only when multiplying the null speed of the asymptotic metric, it is perhaps not surprising that $\eta/s$ is $\beta$-independent, as this is a dimensionless quantity (in the units $\hbar=k_B=1$).

Finally, we note that there is no continuity with the GR results in the $\beta\rightarrow0$ limit when we express quantities in terms of thermodynamic variables. In this limit, the universal horizon of the Ho\v rava solution, which determines the thermodynamic quantities, does not coincide with the spin-2 sound horizon, which determines, for example, $\eta$. In GR, the thermodynamic and sound horizons coincide.

\section{Charge transport from a Ho\v rava black brane}
\label{sec:horavachargecalculation}

We will now address the transport of a conserved $U(1)$ charge in the field theory. Holographically, the situation is identical to the relativistic case, e.g. \cite{Policastro:2002se}.  The source of a $U(1)$ charge and current is a $U(1)$ background potential. These field theory sources are simply dual to the leading near-boundary term (which is $\sim r^0$ for our black brane) of a bulk $U(1)$ gauge potential: the bulk gauge transformations that preserve the radial gauge choice $A_r=0$ act on the boundary values exactly as they should for a field theory potential source. The subleading near-boundary term (in this case $\sim r^1$) of the bulk $U(1)$ gauge potential encodes the expectation values of the conserved $U(1)$ charge and current.

A $U(1)$ gauge potential is comprised of two parts: a spatial vector potential $A_I$, and a scalar potential $\Phi$. These transform as $A_I\to A_I-\partial_I\lambda$ and $\Phi\to\Phi-\partial_t\lambda$ under $U(1)$ gauge transformations, where $\lambda$ is the spacetime dependent gauge parameter. In GR these two potentials combine to form a spacetime vector, but under the less restrictive symmetries of Ho\v rava gravity they are separately well-defined geometric objects. The minimal gauge invariant action of these $U(1)$ potentials which is invariant under foliation preserving diffeomorphisms is 
\begin{equation}
S_{HEM}=-\frac{1}{4\mu_0}\int d^4x N\sqrt{G}\left(F^{IJ}F_{IJ}-\frac{2}{c^2N^2}\left(E_I-F_{JI}N^J\right)\left(E^I-F^{JI}N_J\right)\right),
\label{eq:emaction}
\end{equation}
where the magnetic and electric field strengths are $F_{IJ}\equiv \partial_I A_J-\partial_J A_I$ and $E_I\equiv -\partial_I \Phi+\partial_t A_I$, respectively; $c$ is the speed of electromagnetic waves; and $\mu_0$ is the vacuum permeability, which gives the overall normalization of the action. From the point of view of the action \eqref{eq:emaction}, the speed $c$ is a coupling constant.

The combined action of (\ref{eq:haction}) and (\ref{eq:emaction}) still has the black brane solution (\ref{eq:solu}), with $A_I=0$ and $\Phi=0$. This is an uncharged black brane and, via the dictionary just outlined, this corresponds to a field theory state with zero density of the global $U(1)$ charge. We will concentrate on longitudinal perturbations of the gauge potential. These are dual to charge density and longitudinal current density perturbations of the field theory. It is these operators which should exhibit interesting physics in the hydrodynamic limit, as we outlined in Section \ref{sec:hydrosection}.

The equations of motion for the longitudinal linear fluctuations 
\begin{equation}
\delta A_y(r,t,y)\equiv\int d\omega dk e^{-i\omega t+iky}a(r),\;\;\;\;\;\;\;\;\;\;\;\;\;\;\ \delta\Phi(r,t,y)\equiv\int d\omega dk e^{-i\omega t+iky}\phi(r),
\end{equation}
are
\begin{align}
 k^2 \phi(r)+k \omega a(r)-\left(1-\frac{r^3}{r_h^3}\right)\left[\frac{i k\sqrt{1+\beta}r^3}{r_h^3}a'(r)+\left(1-\frac{r^3}{r_h^3}\right)\phi''(r)\right] &=0,\nonumber\\
k\left(\omega r_h^3-3i\sqrt{1+\beta}r^2\right)\phi(r)+\omega\left(\omega r_h^3-3i\sqrt{1+\beta}r^2\right)a(r)& \nonumber\\
-\left(1-\frac{r^3}{r_h^3}\right)\left[2r^2\left(3c^2+i\omega \sqrt{1+\beta}r+3(1+\beta-2c^2)\frac{r^3}{r_h^3}-3(1+\beta-c^2)\frac{r^6}{r_h^6}\right)a'(r)\right.&\nonumber\\
+\left. i k\sqrt{1+\beta}r^3\phi'(r)+(r_h^3-r^3)\left(-c^2+2c^2\frac{r^3}{r_h^3}+(1+\beta-c^2)\frac{r^6}{r_h^6}\right)a''(r)\right] &= 0,\nonumber\\
 k^2\sqrt{1+\beta}r^3\phi(r)+k\omega\sqrt{1+\beta}r^3a(r)\nonumber\\
-i(r_h^3-r^3)\left[k\left(-c^2+2c^2\frac{r^3}{r_h^3}+(1+\beta-c^2)\frac{r^6}{r_h^6}\right)a'(r)-\omega \phi'(r)\right] &= 0,
\label{ap:max}
\end{align}
where we have not explicitly written that both $a(r)$ and $\phi(r)$ are also functions of $\omega$ and $k$, and we have chosen a gauge where $\delta A_r=0$. Of the two second order equations of motion, only one is linearly independent. This is due to the residual $U(1)$ gauge symmetry and can be made manifest by working with the gauge invariant field 
\begin{align}
e(r)\equiv \omega a(r)+k \phi(r). 
\end{align}
This is a multiple of the electric field component $E_y$, and obeys the second order equation of motion
\begin{align}
& \left[\omega (r_h^9 \omega^2 (\omega r_h^3-3 i \sqrt{1+\beta}r^2 ) + k^2 (1+\beta)r^6  (\omega r_h^6   + 3 i \sqrt{1+\beta}r^2(r_h^3-2r^3)))\right. \nonumber\\ 
&+\left.c^4 k^4 (r_h^3-r^3)^4 - c^2 k^2 (r_h^3 - r^3)^2 (\omega (2 r_h^6 \omega- 3 i  \sqrt{1+\beta}r^2(r_h^3+2r^3))+k^2 (1+\beta)r^6)\right] e(r)\nonumber\\
&+2 i r^2 (r_h^3 - r^3) \omega \left[ - \sqrt{1+\beta} r (\omega (r_h^6 \omega- 3 i \sqrt{1+\beta}r^2(r_h^3 -r^3) )+k^2(1+\beta) r^6 )\right.\nonumber\\
&+\left.c^2 (r_h^3 - r^3)^2 ( 3 i \omega+k^2 \sqrt{1+\beta}r)\right] e'(r)\nonumber\\
&+\left(1-\frac{r^3}{r_h^3}\right)^2\left(c^2(r_h^3-r^3)^2-(1+\beta)r^6\right)\left(\omega^2r_h^6-c^2k^2(r_h^3-r^3)^2+(1+\beta)k^2r^6\right)e''(r) = 0.
\label{ap:eome}
\end{align}

To obtain the correlators of the dual field theory operators, we will follow a similar procedure as in the previous section. The equation of motion (\ref{ap:eome}) has singular points both at the universal horizon $r=r_h$ as well as at $r=r_*$ where
\begin{equation}
\label{eq:rstardefinition}
r_*=\frac{r_h}{\left(1+\frac{\sqrt{1+\beta}}{c}\right)^{1/3}}.
\end{equation}
The latter of these is the sound horizon for excitations of speed $c$. Following our logic in the previous section, we expect this to be the relevant horizon for the linear response calculation. Indeed, this is the only singular point of the equation of motion at which ingoing boundary conditions may be imposed, and is also the outermost singular point. After imposing ingoing boundary conditions at $r_*$, we expand the remaining part of the solution, which is regular at $r=r_*$, in a small $\epsilon$ hydrodynamic expansion
\begin{equation}
e(r)=\left(1-\frac{r}{r_*}\right)^{-\frac{i\omega r_h^3}{3\sqrt{1+\beta}r_*^2}}\left(\epsilon e_1(r)+\epsilon^2e_2(r)+\mathcal{O}(\epsilon^3)\right),
\label{eq:hydroem}
\end{equation}
where $\omega\sim\epsilon^2$ and $k\sim\epsilon$, as before.

In terms of the near-boundary expansions of the original fields
\begin{align}
\label{eq:nearbdyexpansionsgaugefield}
\phi(r\rightarrow0)&=\phi_0(\epsilon)+\phi_1(\epsilon)r+\phi_2(\epsilon)r^2+\phi_3(\epsilon)r^3+\ldots,\\
a(r\rightarrow0)&=a_0(\epsilon)+a_1(\epsilon)r+a_2(\epsilon)r^2+a_3(\epsilon)r^3+\ldots,
\end{align}
the lowest order solutions are
\begin{align}
e_1&=k\phi_0(0)+rk\phi_1(0),\\
e_2&=\left[\omega a_0(0)+k\phi_0'(0)\right]+\left[\omega a_1(0)+k\phi_1'(0)\right]r.
\end{align}
Solving the constraint equation near the boundary imposes that
\begin{align}
a_1(0)=0, && a_1'(0)=-\frac{\omega}{c^2k}\phi_1(0).
\end{align}
And finally, demanding regularity of $e_3$ and $e_4$ at the sound horizon fixes
\begin{align}
\phi_1(0)=\frac{ck^2}{-cr_*k^2+i\omega}\phi_0(0), && \phi_1'(0)=\frac{ck}{-cr_*k^2+i\omega}\left(k\phi_0'(0)+\omega a_0(0)\right).
\end{align}

To determine the Green's functions of the dual operators $\rho$ and $j_\rho^y$, we again need to evaluate the action on this solution. Unlike in the previous section, we do not need to add any boundary terms or counterterms to the bulk action (\ref{eq:emaction}). This $r\rightarrow0$ boundary action is finite and has the form
\begin{align}
S^\text{Gauge}_2\Big|_\text{on-shell}=&\, \int d\omega dk\,\Big[ \phi_0(-\omega,-k)\mathcal{G}_{\phi\phi}(\omega,k)\phi_0(\omega,k)+\phi_0(-\omega,-k)\mathcal{G}_{\phi a}(\omega,k)a_0(\omega,k) \nonumber\\
&+a_{0}(-\omega,-k)\mathcal{G}_{a\phi}(\omega,k)\phi_0(\omega,k)+a_{0}(-\omega,-k)\mathcal{G}_{aa}(\omega,k)a_{0}(\omega,k) \Big ].
\end{align}

Identifying $\phi_0$ and $a_0$ as the sources of $\rho$ and $j_\rho^y$, respectively, we use the prescription (\ref{eq:gfprescription}) to obtain the Green's functions
\begin{align}
\label{eq:holoemresults}
G^R_{q,q}(\omega,k)&=\frac{k^2}{\mu_0c\left(i \omega -c k^2r_*\right)},\nonumber\\
G^R_{q,j_y}(\omega,k)&=\frac{\omega k}{\mu_0c\left(i \omega -c k^2r_*\right)},\nonumber\\
G^R_{j_y,q}(\omega,k)&=G^R_{q,j_y}(\omega,k),\nonumber\\
G^R_{j_y,j_y}(\omega,k)&=\frac{\omega^2}{\mu_0c\left(i \omega -c k^2r_*\right)},
\end{align}
in the hydrodynamic limit. These Green's functions have the characteristic form (\ref{eq:hydrogreensfunctions}) and (\ref{eq:otherhydrogreensfunctions}) associated with the diffusive transport of a conserved charge density. In this case, it is the $U(1)$ charge density.

By comparing the hydrodynamic results to our holographic results (\ref{eq:holoemresults}), we can extract the following expressions for the $U(1)$ charge susceptibility $\chi_\rho$ and diffusion constant $D_\rho$ in the field theory state dual to the black brane solution of Ho\v rava gravity
\begin{equation}
\label{eq:firstemholoresults}
\chi_\rho=\frac{1}{\mu_0 r_*c^2}=\frac{\left(1+\frac{\sqrt{1+\beta}}{c}\right)^{1/3}}{\mu_0 r_hc^2},\;\;\;\;\;\;\;\;\;\;\;\;\;\;\;\;\;\;\;\; D_\rho=cr_*=\frac{cr_h}{\left(1+\frac{\sqrt{1+\beta}}{c}\right)^{1/3}}.
\end{equation}
From the Einstein relation, we can then extract the electrical conductivity
\begin{equation}
\label{eq:holoconductivity}
\sigma=\frac{1}{\mu_0 c}.
\end{equation}
Note that the bulk electromagnetic wave speed $c$ and the corresponding sound horizon radius $r_*$ appear naturally in these formulae. As previously advertised, the diffusion constant obeys the relation (\ref{eq:introdiffusionconstants}). 

As a function of the temperature, the diffusion constant depends on the speeds of both bulk excitations
\begin{equation}
\label{eq:U1diffusionconstT}
D_\rho=\frac{3c\sqrt{1+\beta}}{\left(1+\frac{\sqrt{1+\beta}}{c}\right)^{1/3}}\frac{1}{2\pi T},
\end{equation}
while the dimensionless ratio of momentum and charge diffusion constants depends on their ratio
\begin{equation}
\frac{D_\rho}{D_\pi}=3\cdot 2^{1/3}\frac{c}{\sqrt{1+\beta}}\left(1+\frac{\sqrt{1+\beta}}{c}\right)^{-1/3}.
\end{equation}
When the graviton speed is equal to the speed of light ($c=s_2$), this reduces to the relativistic result \cite{Herzog:2002fn}.

\section{Covariant formalism and field redefinitions}
\label{sec:covarianthorava}

Until this point, we have worked solely with the formalism of Ho\v rava gravity. To better understand our results, and to give further justification for our choice of boundary conditions and holographic renormalization counterterms, it is instructive to express Ho\v rava gravity in a generally covariant way in terms of Einstein-Aether theory \cite{Jacobson:2000xp,Blas:2010hb,Germani:2009yt}. This is a theory of a spacetime metric coupled to a dynamical, timelike `aether' vector $u_X$ of unit norm. The two-derivative action of Einstein-Aether theory is \cite{Janiszewski:2014iaa}
\begin{align}
S^{AE}=\frac{1}{16 \pi G_{AE}}\int d^4 x \sqrt{-g}\,&\biggr( \tilde{R}-2\Lambda-c_1\left(\tilde{\nabla}_Xu^Y\right)\left(\tilde{\nabla}_Xu^Y\right)-c_2\left(\tilde{\nabla}_Xu^X\right)^2 \nn
&~~ -c_3\tilde{\nabla}_Xu^Y\tilde{\nabla}_Yu^X+c_4u^X\tilde{\nabla}_X u^Yu^Z\tilde{\nabla}_Zu_Y\biggr),
\label{eq:kaction}
\end{align}
where $g$ is the determinant of the full spacetime metric $g_{XY}$, $\tilde{\nabla}$ is its covariant derivative, and $\tilde{R}$ is its Ricci scalar. This action is invariant under the full coordinate diffeomorphism symmetry of general relativity. For a hypersurface orthogonal $u_X$ (like that arising in Ho\v rava gravity), not all aether terms in the action are linearly independent, and one of them may be set to zero without loss of generality \cite{Barausse:2011pu}. We choose to set $c_1$=0.

To map onto Ho\v rava gravity, one performs partial gauge fixing by choosing the time coordinate such that $u_X=-N\delta ^t_X$. This is possible when $u_X$ is hypersurface orthogonal. The action (\ref{eq:kaction}) is then equal to the Ho\v rava action, as long as the spacetime metric is decomposed in the ADM form (\ref{eq:adm}). The coupling constants in each action are related in the following way
\begin{align}
\frac{G_H}{G_{AE}}=1+\beta=\frac{1}{1-c_3},&& 1+\lambda=\frac{1+c_2}{1-c_3},&&\alpha=\frac{c_4}{1-c_3}.
\label{eq:consts}
\end{align}
We have been studying the $\alpha=0$ sector of Ho\v rava gravity, which corresponds to $c_4=0$.

\subsection{Momentum transport}
\label{subsec:covariantmapping}

We will now exploit this mapping to relate our Ho\v rava gravity calculations to a more conventional holographic calculation: linear perturbations around the Schwarzschild-AdS$_4$ black brane solution of the Einstein-Hilbert action with a cosmological constant. The metric of this solution is
\begin{equation}
\label{eq:rescaledschwmetric}
d\tilde{s}^2=-\frac{(1+\beta)\left(1-r^3/r_s^3\right)}{r^2}d\tilde{t}^2+\frac{dr^2}{r^2\left(1-r^3/r_s^3\right)}+\frac{1}{r^2}d\vec{x}^2,
\end{equation}
where we have chosen the null speed of the boundary metric at $r=0$ to be $\sqrt{1+\beta}$. The black brane Killing horizon is at $r=r_s$. For reasons that will shortly become clear, we emphasize that this is as a solution to Einstein-Aether theory with $c_1=c_2=c_3=c_4=0$, and with aether vector given by
\begin{align}
\tilde{u}_t=-\sqrt{1+\beta}\frac{1}{r}\left(1-\frac{r^3}{2r_s^3}\right),&& \tilde{u}_r=-\frac{r^2}{2r_s^3}\frac{1}{\left(1-\frac{r^3}{r_s^3}\right)}.
\end{align}
By a temporal diffeomorphism, we may transform this Schwarzschild black brane into the unfamiliar form
\begin{equation}
d\hat{s}^2=-\frac{(1+\beta)(r_s^3-r^3)}{r^2r_s^3}d\hat{t}^2+2\frac{\sqrt{1+\beta}r}{2r_s^3-r^3}drd\hat{t}+\frac{4r_s^6}{r^2(2r_s^3-r^3)^2}dr^2+\frac{1}{r^2}d\vec{x}^2,
\label{eq:ghat}
\end{equation}
with
\begin{align}
\hat{u}_t=-\sqrt{1+\beta}\frac{1}{r}\left(1-\frac{r^3}{2r_s^3}\right),&&\hat{u}_r=0.
\end{align}
We now perform a non-trivial field redefinition
\begin{align}
g_{XY}\equiv\hat{g}_{XY}+\frac{(\sigma-1)}{\sigma}\hat{u}_X\hat{u}_Y,&& u_X\equiv \hat{u}_X/\sqrt{\sigma},
\label{eq:redef}
\end{align}
on the metric and the aether field, where we have chosen the constant $\sigma=s_2^2=1+\beta$. After this field redefinition, the solution takes the form
\begin{align}
\label{eq:finalmetricsol}
&ds^2=\left[-\frac{1}{r^2}\left(1-\frac{2r^3}{r_h^3}\right)+\frac{\beta r^4}{r_h^4}\right]dt^2+2\frac{\sqrt{1+\beta}r}{r_h^3-r^3}drdt+\frac{r_h^6}{r^2(r_h^3-r^3)^2}dr^2+\frac{1}{r^2}d\vec{x}^2,\\
&u_t=-\frac{1}{r}\left(1-\frac{r^3}{r_h^3}\right),
\end{align}
where we have relabelled $r_h=2^{1/3}r_s$.

This solution is identical to the Einstein-Aether formulation of our Ho\v rava gravity solution (\ref{eq:solu}). This is not a coincidence. Under the field redefinitions (\ref{eq:redef}),  the Einstein-Aether action maps to itself, but with $G_{AE}\rightarrow G_{AE}/\sqrt{\sigma}$ and with different values of the coupling constants $c_i$ \cite{Foster:2005ec}. As we began with all $c_i=0$, the resulting action has non-zero $c_2=-c_3$ and $c_1=c_4=0$, taking into account hypersurface orthogonality. Translating this hypersurface orthogonal Einstein-Aether theory to Ho\v rava gravity using (\ref{eq:consts}), we conclude that the net effect of these coordinate transformations and field redefinitions is to convert the GR action into the Ho\v rava gravity action with $\beta\ne0$, $\lambda=\alpha=0$. 

This explains why the series of coordinate transformations and field redefinitions we explicitly performed above turns the Schwarzschild-AdS solution into our Ho\v rava gravity solution. General solutions of the Ho\v rava action with $(\alpha,\lambda)\ne0$ cannot be mapped to GR solutions, but since the Ho\v rava solution has $\alpha=0$ and is independent of $\lambda$, we can set $\lambda=0$ and utilize the mapping above.

The mapping is much more powerful than this. Not only is our background solution of Ho\v rava gravity independent of $\lambda$, but the equations of motion for linearized perturbations of the transverse gravitons around this state are also independent of $\lambda$. This implies that, after suitable field redefinitions and coordinate transformations, the GR equations of motion for linearized transverse perturbations around the Schwarzschild-AdS solution are equivalent to the linearized Ho\v rava equations of motion (\ref{eq:eomsshear}) around the solution (\ref{eq:solu}). To be explicit, they are identical after identifying $\delta h_{XY}$, the perturbations of the Schwarzschild-AdS spacetime (\ref{eq:rescaledschwmetric}), with
\begin{equation}
\begin{aligned}
&\delta h_{yx}=\delta G_{yx},\\
&\delta h_{\tilde{t}x}=\delta N_x,\\
&\delta h_{rx}=\delta G_{rx}+\frac{r^3r_s^3}{\sqrt{1+\beta}\left(r_s^3-r^3\right)\left(2r_s^3-r^3\right)}\delta N_x,
\end{aligned}
\end{equation}
and then replacing
\begin{equation}
\begin{aligned}
\label{eq:coordinatechange}
&\frac{\partial}{\partial\tilde{t}}\rightarrow\frac{\partial}{\partial t},\\
&\frac{\partial}{\partial r}\rightarrow\frac{\partial}{\partial r}+\frac{r^3r_s^3}{\sqrt{1+\beta}\left(r_s^3-r^3\right)\left(2r_s^3-r^3\right)}\frac{\partial}{\partial t}.
\end{aligned}
\end{equation}

As the Schwarzschild-AdS calculation is very well-understood, we can use this equivalence to shed light on the results, and the calculational details, of our Ho\v rava gravity computations in Section \ref{sec:horavamomentumcalculation}. Firstly, we note that the black brane Killing horizon of the Schwarzschild solution maps to the spin-2 sound horizon $r_s$ of the Ho\v rava gravity solution. Therefore, our imposition of ingoing boundary conditions at this sound horizon in the Ho\v rava formulation of the calculation is equivalent to imposing them at the Killing horizon of the Schwarzschild metric. Secondly, the counter-terms (\ref{eq:hcts}) we added to obtain a finite on-shell Ho\v rava action are equivalent to those of the relativistic case \cite{deHaro:2000vlm} after the appropriate transformations. These give justification for our choice of boundary conditions and counter-terms in the Ho\v rava formulation of the problem.

Under the mapping described above, the coefficients $n_0,n_3,g_0,g_3$ in the near-boundary expansions (\ref{eq:nearbdyexpansions}) of the Ho\v rava gravity fields map in a trivial way to the corresponding coefficients of the near-boundary expansions of the fields $\delta h_{xt},\delta h_{xy}$ in the Schwarzschild-AdS calculation. There should therefore be a close relationship between the Green's functions obtained from Ho\v rava gravity and those obtained from the Schwarzschild-AdS black brane in \cite{Herzog:2002fn}. The Green's functions obtained from the Ho\v rava calculation should be those of the relativistic case, in which the black brane Killing horizon radius is replaced by the spin-2 sound horizon radius, and the relativistic speed of light is replaced by the spin-2 speed $s_2=\sqrt{1+\beta}$. The latter of these conditions is because, after the mapping, the asymptotic null speed of the Schwarzschild-AdS solution (\ref{eq:rescaledschwmetric}) is given by $\sqrt{1+\beta}$. This is precisely what we found in Section \ref{sec:momentumtransportGFresults}.

Finally, we note that we are not aware of any way of mapping thermodynamic quantities using this procedure. Although the Schwarzschild-AdS black brane Killing horizon maps to the spin-2 horizon of the Ho\v rava gravity solution, this is not the causal horizon of the theory, as there may be excitations that travel at speeds greater than $s_2$. 
This means that there is no direct way of mapping thermodynamic quantities associated with the horizon, such as the entropy and temperature, between the solutions. 
A covariant expression of the location of the causal horizon in Einstein-Aether theory is given by $u_X\chi^X=0$, where $\chi^X$ is the asymptotic time-like Killing vector. The universal horizon is at $r=r_h$ for both $g_{XY}$ and $\hat{g}_{XY}$ above \cite{Berglund:2012bu,Berglund:2012fk,Bhattacharyya:2014kta}. 

\subsection{Charge transport}

It is also possible to recast our calculation of charge diffusion in Section \ref{sec:horavachargecalculation} in terms of a covariant Einstein-Aether theory. The electromagnetic action \eqref{eq:emaction} can be written in the Einstein-Aether formalism as \cite{Balakin:2014tza}
\begin{align}\label{SEAM}
S^{EAM}  =  -\frac{1}{4\mu_0} \int d^4 x \sqrt{-g} \left(F_{XY} F^{XY} - \kappa u_X F^{XY} u_Z F^Z_{~Y} \right),
\end{align}
where $F_{XY} = \tilde\nabla_X A_Y - \tilde\nabla_Y A_X$. In terms of the coupling constants used in (\ref{SEAM}), the speed of electromagnetic waves is
\begin{align}
c^2=\frac{2}{2+\kappa}.
\end{align}
The action \eqref{SEAM} differs from the usual covariant Maxwell action due to the second term proportional to $\kappa$. We can again take advantage of field redefinitions and coordinate transformations to make this situation more intuitive. Under the field redefinitions (where $\sigma$ is a constant)
\begin{align}
\hat{g}_{AB}=g_{AB}-\left(\sigma-1\right)u_Au_B,&& \hat{u}_A=u_A\sqrt{\sigma},&&\hat{F}^{AB}=F^{AB},
\end{align}
the action (\ref{SEAM}) transforms to itself, but with different values of the coupling constants $\kappa$ and $\mu_0$. This field redefinition is the inverse of that in (\ref{eq:redef}).

By choosing $\sigma$ to be the speed of light squared
\begin{equation}
\sigma=c^2,
\end{equation}
the new metric of the black brane solution is
\begin{equation}
d\hat{s}^2=-\frac{c^2\left(1-2\frac{r^3}{r_h^3}+\left(1-\frac{1+\beta}{c^2}\right)\frac{r^6}{r_h^6}\right)}{r^2}d\hat{t}^2+\frac{2r\sqrt{1+\beta}}{r_h^3-r^3}d\hat{t}dr+\frac{1}{r^2}d\vec{x}^2+\frac{dr^2}{r^2\left(1-\frac{r^3}{r_h^3}\right)^2},
\end{equation}
which can be brought into the nicer diagonal form
\begin{equation}
\begin{aligned}
\label{eq:transformedmetricmaxwell}
&d\tilde{s}^2=-\frac{c^2f(r)}{r^2}d\tilde{t}^2+\frac{1}{r^2}d\vec{x}^2+\frac{dr^2}{r^2f(r)},\\
&f(r)=1-2\frac{r^3}{r_h^3}+\left(1-\frac{1+\beta}{c^2}\right)\frac{r^6}{r_h^6},
\end{aligned}
\end{equation}
with a radially dependent temporal diffeomorphism. This new metric is an ``effective metric'' for the electromagnetic modes, meaning that the sound horizon for modes of speed $c$ in the Ho\v rava theory is now a Killing horizon.\footnote{For $c_3 = (1-c^2+\beta)/(1+\beta)$ and adequate $u_t$ and $u_r$, equation \eqref{eq:transformedmetricmaxwell} is a solution. It would be intersting to investigate this solution further.} The null speed in the asymptotic region of the new metric is $c$. Although this is not the Schwarzschild-AdS solution, it gives an intuitive reason for why it is the speed of light $c$ and the sound horizon $r_*$ for light, rather than the null speed of the asymptotic metric and the universal horizon, which appear naturally in the Ho\v rava results (\ref{eq:holoemresults}). In the limit $c^2=1+\beta$, i.e.~when the spin-2 graviton and light both move at the same speed, this reduces to the Schwarzschild-AdS solution. After the field redefinition, the new coupling constants in the action are
\begin{align}
\hat{\mu}_0=c\mu_0,&&\hat{\kappa}=\frac{2}{c^4}\left(1-c^4\right),
\end{align}
and so the action of the redefined fields is still not the Maxwell action.

\subsection{Sound waves when $\lambda=0$}

Finally, we will comment briefly on longitudinal linearized perturbations around our Ho\v rava solution. This will tell us, amongst other things, about the longitudinal transport of momentum in the dual field theory. There is one limit in which these transport properties are relatively simple to deduce.

In Section \ref{subsec:covariantmapping}, we described how to relate our $\lambda=0$ Ho\v rava equations of motion to those of perturbations around the Schwarzschild-AdS solution of GR. For the transverse linearized perturbations of Ho\v rava gravity, the equations of motion are independent of $\lambda$ and thus we could perform this mapping for any $\lambda$. The longitudinal perturbation equations are $\lambda$-dependent and so generically we cannot use this mapping. However, in the special case of $\lambda=0$, we can perform the mapping without any problems. Explicitly, the equations of motion for perturbations $\delta h_{XY}$ of the Schwarzschild-AdS metric (\ref{eq:rescaledschwmetric}) are transformed into the Ho\v rava equations by identifying
\begin{equation}
\begin{aligned}
&\delta h_{\tilde{t}\tilde{t}}=-2\left(1+\beta\right)N\delta N-\left(\frac{N_r}{G_{rr}}\right)^2\delta G_{rr}+\frac{2N_r}{G_{rr}}\delta N_r,\\
&\delta h_{xx}=\delta G_{xx},\\
&\delta h_{yy}=\delta G_{yy},\\
&\delta h_{rr}=\delta G_{rr}+2F(r)\delta N_r+F(r)^2\left(-2\left(1+\beta\right)N\delta N-\left(\frac{N_r}{G_{rr}}\right)^2\delta G_{rr}+\frac{2N_r}{G_{rr}}\delta N_r\right),\\
&\delta h_{r\tilde{t}}=\delta N_r+F(r)\left(-2\left(1+\beta\right)N\delta N-\left(\frac{N_r}{G_{rr}}\right)^2\delta G_{rr}+\frac{2N_r}{G_{rr}}\delta N_r\right),\\
&\delta h_{y\tilde{t}}=\delta N_y,\\
&\delta h_{yr}=\delta G_{yr}+F(r)\delta N_y,
\end{aligned}
\label{eq:perttrans}
\end{equation}
where
\begin{equation}
F(r)=\frac{r^3r_s^3}{\sqrt{1+\beta}}\frac{1}{\left(r_s^3-r^3\right)\left(2r_s^3-r^3\right)},
\end{equation}
followed by the replacement (\ref{eq:coordinatechange}). We have not written either set of longitudinal equations explicitly as they are very lengthy.

As the near-boundary expansions of each field map in a trivial way, we can painlessly determine the longitudinal, hydrodynamic quasi-normal modes of the Ho\v rava solution (\ref{eq:solu}), where we define a quasi-normal mode as a solution that is ingoing at the sound horizon and whose leading term vanishes at the asymptotic boundary. By replacing the Killing horizon radius with the spin-2 horizon radius, and the asymptotic null speed with $\sqrt{1+\beta}$, in the Schwarzschild-AdS results \cite{Herzog:2003ke}, we find the dispersion relations
\begin{equation}
\label{eq:sounddispersionrelation}
\omega=\pm\sqrt{\frac{1+\beta}{2}}k-i\frac{\sqrt{1+\beta}\,r_s}{6}k^2.
\end{equation}
These are the dispersion relations of the sound waves in the field theory dual of Ho\v rava gravity with $\alpha=\lambda=0$. Once again, we note that the speed and the attenuation coefficient are most naturally expressed in terms of the speed of the spin-2 graviton, rather than the null speed of the boundary metric. The attenuation coefficient may be rewritten as $\Gamma=(1+\beta)\eta/2sT=D_\pi/2$ \cite{Herzog:2003ke}. In the more general case of $\lambda\ne 0$ we expect the situation to be more complicated, as there will be another graviton excitation.

In \cite{Eling:2014saa}, the dual hydrodynamics of the solution (\ref{eq:solu}) was studied (perturbatively in $\beta$) within Einstein-Aether theory and found to be consistent with relativistic hydrodynamics to first order in the derivative expansion. In our Horava theory calculation, this is only the case when $\lambda=0$. This apparent discrepancy is because the boosted solutions of \cite{Eling:2014saa} are not solutions of Horava gravity in a global time.

\section{Discussion}
\label{sec:discussion}

We have used the non-relativistic holographic duality conjectured in \cite{Janiszewski:2012nb,Janiszewski:2012nf}, and reinforced in the language of \cite{Jensen:2014aia,Hartong:2015zia}, to study the collective transport properties of a field theory dual to a black brane solution of Ho\v rava gravity. 
In agreement with the general principles of hydrodynamics outlined in Section \ref{sec:hydrosection}, we have shown that both charge and transverse momentum diffuse, and calculated the diffusion constants and conductivities associated with these processes, in equations (\ref{eq:holodiffconstant}), (\ref{eq:holoviscosity}), (\ref{eq:firstemholoresults}) and (\ref{eq:holoconductivity}). The ratio (\ref{eq:etaovershorava}) of entropy density to viscosity is independent of $\beta$ and is increased by a factor of $2^{2/3}$ compared to the relativistic case, in agreement with the conjecture of \cite{Eling:2014saa}. Geometrically, this factor stems from the fact that the thermodynamic horizon does not coincide with the relevant trapped surface, i.e.~the spin-2 sound horizon. From the point of view of the gravitational theory, we have derived new results for the hydrodynamic quasi-normal modes of the solution (\ref{eq:solu}) of Ho\v rava gravity.

We have outlined how the canonical method for determining linear response properties from a dual theory of relativistic gravity should be modified for the non-relativistic case of Ho\v rava gravity. The major difference is that, in Ho\v rava gravity, different excitations (e.g.~different gravitons, the photon) have different speeds, and these speeds are different from the null speed of the asymptotic metric. In GR, Lorentz invariance fixes these speeds to all be the same. In Ho\v rava gravity, these different excitations each have their own independent sound horizon, which specifies the region from which the excitation cannot escape. In principle, these are independent from the universal horizon, the causal horizon of the spacetime itself. We propose that calculation of the two point retarded Green's function of the dual field theory requires one to impose ingoing boundary conditions at the sound horizon of the dual excitation. 

We have focused here on the most tractable Ho\v rava gravity calculations: the diffusive excitations around the analytic black brane solution (\ref{eq:solu}). These examples make it easiest to identify the fundamental differences with the relativistic calculations of two-point functions, and to identify how the canonical methods should be modified. Indeed, in some of these cases we had the luxury of using the mapping to GR described in Section \ref{sec:covarianthorava} to check that our proposed modifications are sensible. In general, such a simple mapping is not possible and one should then directly use the procedure we have outlined for Ho\v rava gravity.

There is one important feature of Ho\v rava gravity that does not affect the examples we have studied here: the existence of new graviton excitations due to the reduced diffeomorphism symmetry. These appear to be present in the $\lambda\ne0$ calculation of longitudinal transport, leading to a complicated set of coupled equations for linearized excitations, in contrast to the single equation when $\lambda=0$. The most immediate extension of our work is therefore to complete our understanding of the dynamics in the longitudinal sector, and to identify the nature and consequences of this new degree of freedom in the field theory. Due to the existence of this mode, we expect the hydrodynamics of the longitudinal sector to be significantly different from that of the relativistic case.

A second important generalization of this work is to study Ho\v rava theories with $\alpha\ne0$. These are spacetimes whose asymptotic boundary metric has a scaling exponent $z\ne1$, which is not unusual in the low energy limit of many-body systems. Given the recent holography-inspired success of applying hydrodynamic techniques to explain the transport properties of relativistic many-body systems, we are hopeful that the non-relativistic generalization will be similarly useful. For example, in relativistic hydrodynamics, energy cannot flow independently of momentum, and thus the thermoelectric conductivity matrix depends only on one transport coefficient \cite{Davison:2015taa}. This will not generically be the case for a system without Lorentz symmetry, and it is worthwhile verifying this directly from the dual gravitational perspective. As mentioned in the dictionary discussion of Section \ref{sec:pert}, 
the Ho\v rava gravity theory~\eqref{eq:haction} 
seems unable to encode the source of the energy current, as there is no bulk field corresponding to the spatial components $n_i$ of the Newton-Cartan clock vector. To restore this source and allow calculation of correlators of the energy current 
will require the study of a more general form of Ho\v rava gravity,
a task explored in \cite{Hartong:2015zia}. There is undoubtedly fruitful work to be pursued in this direction.

Although we have determined the counter-terms necessary to holographically renormalize the Ho\v rava action for the diffusive channel, a complete description of the process is still lacking, although progress was made in \cite{Griffin:2012qx}. A reasonable guiding principle is the analog of the GR procedure: counter-terms should be constructed out of geometrically invariant boundary data. For Ho\v rava gravity, this means terms invariant under foliation preserving diffeomorphisms. In GR, by choosing radial gauge $g_{r\mu}=0$ the only boundary data is the induced metric $g_{\mu\nu}|_{r=0}$; in Ho\v rava gravity, due to the restricted symmetries, radial gauge sets $G_{ri}=0$ for the spatial metric, but the radial component of the shift vector $N_r$ generically does not vanish. Therefore, in addition to the boundary data of the induced spatial metric $G_{ij}$, the induced shift vector $N_i$, and lapse $N$, the boundary value of $N_r$ should be included in possible counter-terms. 

In the relativistic case, the calculation of the dual linear response conductivities can be greatly simplified using `membrane paradigm' techniques, that relate them to properties of the black brane horizon \cite{Iqbal:2008by,Davison:2015taa,Gursoy:2014boa,Grozdanov:2016ala}. It may be possible to generalize these techniques to Ho\v rava gravity, in which case we expect the properties of the sound horizon to play the corresponding role. A more challenging task is to determine the full non-linear response of the black brane and interpret it in terms of the non-linear hydrodynamics of a dual theory. Some work in this direction was undertaken in \cite{Eling:2014saa}.

Finally, it would be interesting to construct and study charged black brane solutions of Ho\v rava gravity. Except in the limit $s_2=c=1$, It does not appear to be possible to reverse engineer a charged solution of Ho\v rava gravity by starting with the Reissner-Nordstrom-AdS solution of GR and using the mappings described in Section \ref{sec:covarianthorava}.

\paragraph*{Acknowledgements} We are very grateful to Steffen Klug for his collaboration during the initial stages of this project. We thank Christopher Eling and Jelle Hartong for useful discussions and remarks on the draft of this paper. The work of R.A.D.~is supported by the Gordon and Betty Moore Foundation EPiQS Initiative through Grant GBMF\#4306. S.G.~is supported in part by a VICI grant of the Netherlands Organisation for Scientific Research (NWO), and by the Netherlands Organisation for Scientific Research/Ministry of Science and Education (NWO/OCW). The work of S.J.~is supported in part by NSERC of Canada. M.K.~thanks the Instituut-Lorentz and Koenraad Schaalm for hospitality during important phases of this work.

\bibliographystyle{JHEP}
\bibliography{shear-horava}

\end{document}